\documentclass[prl,reprint,superscriptaddress,showpacs,preprintnumbers,nobibnotes,amsmath,amssymb,aps,longbibliography]{revtex4-1}

\usepackage{cancel}
\usepackage{accents}
\usepackage{mciteplus,slashed}
\usepackage{amssymb,cancel,amsmath}
\usepackage{dcolumn}
\usepackage{bm}
\usepackage[caption=false]{subfig}
\usepackage{appendix}
\usepackage{feynmp-auto}
\usepackage[T1]{fontenc}	
\usepackage{csvsimple}
\usepackage[colorlinks=true,citecolor=blue,linkcolor=blue]{hyperref}
\usepackage[section]{placeins}
\usepackage[capitalise]{cleveref}
\usepackage{booktabs}
\usepackage{graphicx}
\usepackage{subfig}
\usepackage{mathrsfs}
\usepackage{multirow}
\usepackage[utf8]{inputenc}
\usepackage{siunitx}
\usepackage{comment}
\usepackage{gensymb}
\usepackage{enumitem}
\usepackage{wrapfig}

\usepackage[dvipsnames]{xcolor}
\usepackage[normalem]{ulem}

\newcommand{\SIREN}{\texttt{SIREN}}

\begin{document}




\title{Lake- and Surface-Based Detectors for Forward Neutrino Physics}

\author{Nicholas W. Kamp}
\email{nkamp@g.harvard.edu}
\affiliation{Department of Physics and Laboratory for Particle Physics and Cosmology, Harvard University, Cambridge, MA 02138, US}

\author{Carlos A. Arg\"{u}elles}
\email{carguelles@g.harvard.edu}
\affiliation{Department of Physics and Laboratory for Particle Physics and Cosmology, Harvard University, Cambridge, MA 02138, US}

\author{Albrecht Karle}
\email{akarle@wisc.edu}
\affiliation{Department of Physics and Wisconsin IceCube Particle Astrophysics Center, University of Wisconsin– Madison, Madison, WI 53706, USA}

\author{Jennifer Thomas}
\email{jennifer.thomas@ucl.ac.uk}
\affiliation{Department of Physics and Wisconsin IceCube Particle Astrophysics Center, University of Wisconsin– Madison, Madison, WI 53706, USA}
\affiliation{Department of Physics and Astronomy, University College London, London, WC1E 6BT, UK}

\author{Tianlu Yuan}
\email{tyuan@icecube.wisc.edu}
\affiliation{Department of Physics and Wisconsin IceCube Particle Astrophysics Center, University of Wisconsin– Madison, Madison, WI 53706, USA}

\date{\today}

\begin{abstract}
We propose two medium-baseline, kiloton-scale neutrino experiments to study neutrinos from LHC proton-proton collisions: SINE, a surface-based scintillator panel detector observing muon neutrinos from the CMS interaction point, and UNDINE, a water Cherenkov detector submerged in lake Geneva observing all-flavor neutrinos from LHCb.
Using a Monte Carlo simulation, we estimate millions of neutrino interactions during the high-luminosity LHC era.
We show that these datasets can constrain neutrino cross sections, charm production in $pp$ collisions, strangeness enhancement as a solution to the cosmic-ray muon puzzle, and heavy neutral leptons.
SINE and UNDINE thus offer a cost-effective medium-baseline complement to the proposed short-baseline forward physics facility.
\end{abstract}

\maketitle


\section{Introduction}

\begin{figure*}[ht]
    \centering
    \includegraphics[width=\linewidth]{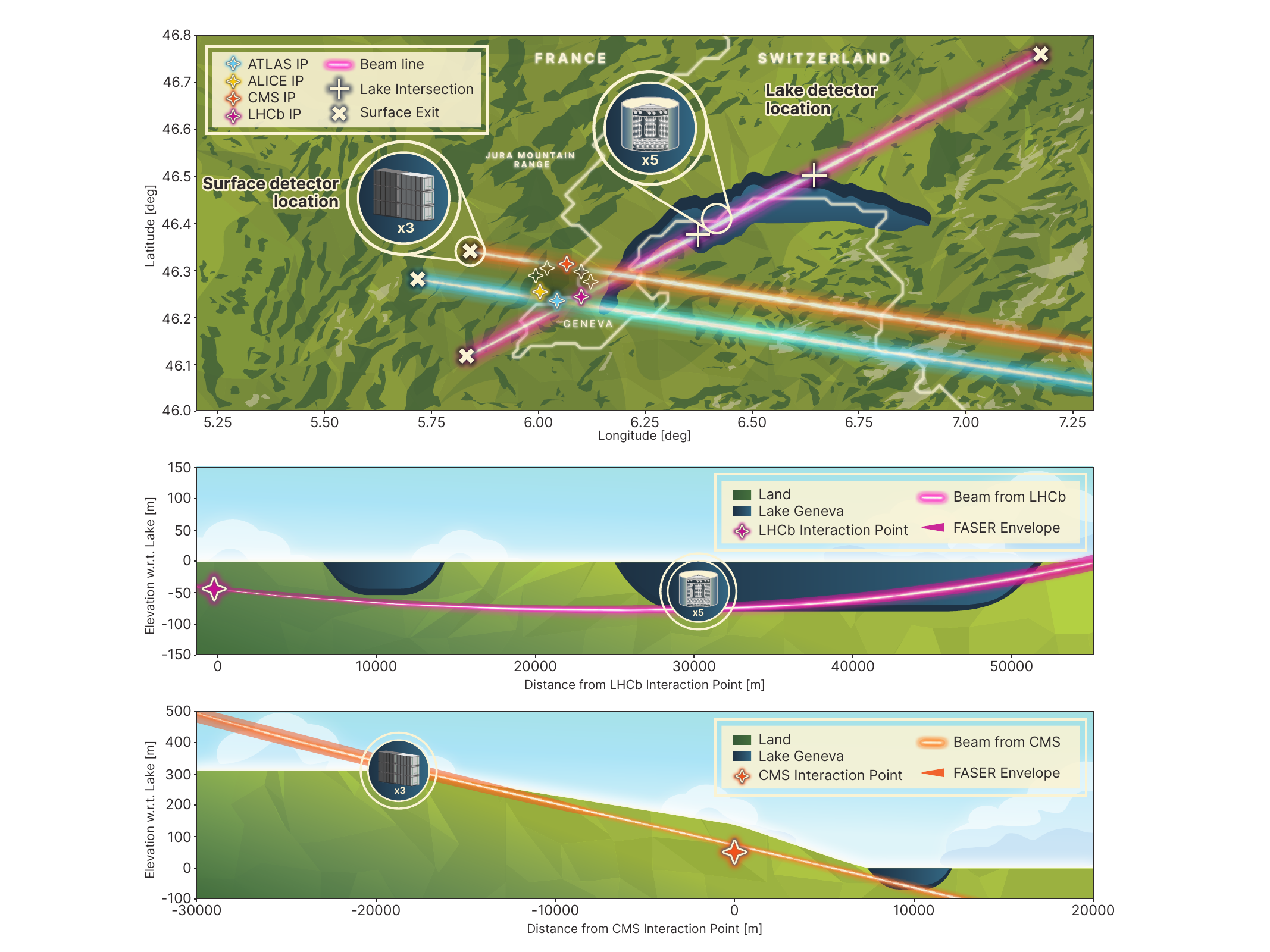}
    \caption{\textbf{\textit{Illustration of experimental layout.}} Top: shows sky-view vision of the area around the LHC including potential locations for the detectors. Middle and bottom: shows a side view of the UNDINE and SINE detector setups.
    The shaded envelope around each beamline corresponds to the profile of the neutrino beam that passes through the FASER detector.
    The geometries shown in this figure are only approximate and for illustration purposes only. Illustration by Jackapan Pairin.}
    \label{fig:geometry-diagram}
\end{figure*}

It has been realized since 1984~\cite{DeRujula:1984pg} that proton-proton ($pp$) collisions at the large hadron collider (LHC) produce a collimated beam of neutrinos along the collision axis, or the ``forward direction''~\cite{DeRujula:1984pg,DERUJULA199380,FERNANDEZ1993326}.
These neutrinos come from the decay of forward-going hadrons, including pions, kaons, charm mesons, and B mesons~\cite{DeRujula:1984pg,DERUJULA199380}.
They offer a unique opportunity to study the interaction of TeV neutrinos in a controlled laboratory setting, enabling precise measurements of the neutrino cross section at these energies~\cite{FASER:2024hoe}.
Forward neutrinos also carry important hadronic information in both their production and their detection~\cite{Feng:2022inv}.
Furthermore, a relatively large number of tau neutrinos are produced, mostly from $D$ and $D_s$ decays~\cite{DERUJULA199380}.
Observations of these tau neutrinos could substantially enlarge the global collection rate, enabling unique physics~\cite{MammenAbraham:2022xoc}.

Collider-generated neutrinos from the LHC were observed for the first time in 2023 by the \textsc{ForwArd SEaRch} (FASER) experiment~\cite{FASER:2023zcr}.
Since then, FASER has performed the first measurements of the total $\nu_e$ and $\nu_\mu$ cross sections in the TeV energy regime~\cite{FASER:2024hoe}.
Following the success of FASER, there are plans to construct the \textsc{Forward Physics Facility} (FPF)~\cite{Anchordoqui:2021ghd,Adhikary:2024nlv} by excavating a cavern underground along the ATLAS forward beamline with enough space to house larger detectors.
These larger detectors have various physics goals: searches for the decay of new, heavy particles --- FASER2~\cite{Feng:2022inv} ---, neutrino and light dark matter detection --- FASER$\nu2$~\cite{Batell:2021aja,Feng:2022inv}, Advanced SND@LHC~\cite{Abbaneo:2895224}, FLArE~\cite{Batell:2021blf},--- and searches for millicharged particles --- FORMOSA~\cite{Foroughi-Abari:2020qar}.
Beyond the objectives listed above, the FPF will also have world-leading sensitivity to many models of hypothetical long-lived particles produced along the forward direction in $pp$ collisions~\cite{Anchordoqui:2021ghd}.

Despite the success and vitality of this program, we would like to point out two important observations relevant to the neutrino physics aspects of the existing program.
First, the neutrino beam produced at the $pp$ collision is very collimated compared to traditional neutrino beams~\cite{MiniBooNE:2008hfu,Adamson:2015dkw,T2K:2019eao}, with a spread of $\mathcal{O}(10\,{\rm m})$ at distances of $\mathcal{O}(10\,{\rm km})$.
This implies that a detector placed far away from the interaction point will still capture most of the beam with a relatively small experiment.
The key advantage is that such a detector would be unconstrained from near-siting logistics and beam-induced backgrounds, such as high-energy muons.
Second, the intensity of the forward neutrino beam of the high-luminosity LHC (HL-LHC) is such that millions of neutrinos can be observed with kiloton-scale detectors. 
This implies that these experiments can achieve a high signal-to-background ratio at the surface level despite cosmic-ray backgrounds.

We build upon these insights as well as prior work on experiments deployed in natural environments--- IceCube~\cite{IceCube:2016zyt}, KM3NeT~\cite{Margiotta:2014gza}, and CHIPS~\cite{CHIPS:2024vpb}---and exploit the local geography around the LHC to bring kiloton-scale detectors to collider neutrino physics.

In this letter, we introduce two new detector concepts for medium-baseline forward neutrino (MBF$\nu$) physics.
The main idea is to deploy kiloton-scale detectors leveraging the surrounding geography at distances of $\mathcal{O}(10\,{\rm km})$ from the LHC interaction points.
The first detector leverages the fact that neutrinos from the CMS interaction point travel through 18\,km of bedrock before exiting Earth's surface.
We propose to use scintillator panels to observe upward-going muons from Earth's surface from upstream neutrino interactions in the preceding bedrock.
We refer to this concept as the Surface-based Integrated Neutrino Experiment (SINE).
The second detector leverages the fact that neutrinos from the LHCb interaction point pass through lake Geneva by submerging kiloton-scale water Cherenkov detectors around 50\,m below the lake's surface.
This is the UNDerwater Integrated Neutrino Experiment (UNDINE)~\footnote{UNDINE derives from the Latin word \textit{unda} for wave and is also the name for a water nymph in European folklore.}.
SINE and UNDINE are cost-effective complements to the planned FPF detectors and can collect $\mathcal{O}(10{\rm M})$ and  $\mathcal{O}(0.1{\rm M})$ neutrino interactions over the course of the HL-LHC run, respectively.
These detectors are represented diagrammatically in \cref{fig:geometry-diagram}.

The remainder of the letter is organized as follows.
\Cref{sec:detectors} describes the SINE and UNDINE detectors in more detail as well as the forward neutrino beamline model used.
Next, \cref{sec:rates} presents a simulation-based calculation of the neutrino interaction rates observed by SINE and UNDINE during the HL-LHC.
We discuss three physics opportunities enabled by these datasets in \cref{sec:physics}: cross sections, charm production, and constraints on a solution to the cosmic-ray muon puzzle.
We conclude in \cref{sec:conclusion} with the next steps toward deploying SINE and UNDINE for the HL-LHC.

\section{The Lake and Surface Detectors \label{sec:detectors}} 

The SINE and UNDINE detectors rely on the forward neutrino flux produced at CMS and LHCb, respectively.
A birds-eye view of the neutrino beam from these two interaction points is shown in the top panel of \cref{fig:geometry-diagram}.
To model this flux, we use the forward neutrino flux computed in ATLAS presented in Ref.~\cite{forward-nu-flux,Kling:2023tgr}.
While there will be differences between the flux profiles at each interaction point, this is a reasonable approximation for this first study.
More details about the flux profile assumed here are given in~\cref{app:flux}.

We now turn to the detectors themselves.
As depicted in the bottom panel of \cref{fig:geometry-diagram}, the SINE detector looks for upward-going muons using scintillator panels located at the point through which the forward neutrino beamline from the CMS interaction point exits the Earth's surface, corresponding to a distance of approximately $18\,{\rm km}$.
At TeV energies, muons will travel upwards of $1\,{\rm km}$ before losing a substantial fraction of their energy~\cite{Koehne:2013gpa}; thus, the muons created in $\nu_\mu$ charged-current (CC) interactions in the bedrock will often lead to muons exiting Earth's surface.
If the scintillator panels are segmented in one or two dimensions, directionality cuts can separate these neutrino-induced muons from cosmic-induced muons.
We study the cosmic-ray background rejection using \textsc{EcoMug}~\cite{Pagano:2021wqu} and find that the combination of beam timing and spatial segmentation will allow for a signal-to-background ratio much greater than one.
This is discussed further in \cref{sec:bkg}.

The nominal SINE design consists of a series of modules, each of which is a standard $12.2\,{\rm m} \times 2.4\,{\rm m} \times 2.6\,{\rm m}$ shipping container with scintillator instrumenting the front and back.
These modules are arranged in sets of two wide by three tall.
We consider three of these $2\times3$ clusters, the first of which is placed such that the CMS neutrino beamline intersects the center of the cluster and the other two are placed $100\,{\rm m}$ in front and behind.
Though the central detector will have the highest rate of up-going muons from neutrinos, the latter two detectors are still effective due to the small angle of the CMS neutrino beam with respect to the surface.
The modular design of the detectors is intended to make the deployment as easy as possible.
The SINE modules do not need to be constructed locally, and it will be straightforward to ship them to the surface exit point.
Furthermore, the modules can be designed to be self-contained by including the data acquisition system, power supplies, and other supporting infrastructure within each shipping crate.

The UNDINE detector, depicted in the middle panel of \cref{fig:geometry-diagram}, looks for neutrino interactions along the LHCb forward beamline using water Cherenkov detectors.
The detector will be situated approximately 30\,km from the LHCb interaction point.
We follow the CHIPS concept~\cite{CHIPS:2024vpb}, which deployed a $5\,{\rm kt}$ water Cherenkov detector in a water-filled mine pit near Hoyt Lakes, Minnesota, along the Fermilab Neutrino Main Injector beamline.
The detector itself will be filled with purified water to maximize light yield, while the surrounding lake water will serve as a natural overburden.
For the original CHIPS detector, $50\,{\rm m}$ of water overburden was sufficient to isolate neutrino interactions from cosmic muons.

The nominal UNDINE design consists of five CHIPS-style modules, each of which is a $12.5\,{\rm m}$ radius,  $12.5\,{\rm m}$ height cylinder instrumented with photo-multiplier tubes (PMTs) along the inner surface.
Thus, the full UNDINE detector corresponds to an instrumented mass of approximately $30\,{\rm kt}$.
The water Cherenkov detectors of UNDINE will have finer reconstruction capabilities compared to SINE.
As demonstrated by Super-Kamiokande, we expect UNDINE to be able to separate electron and muon flavored atmospheric neutrinos up to TeV energies~\cite{Super-Kamiokande:2015qek}.
We also expect an energy resolution of $0.2-0.4$ in $\log_{10}(E_\nu / {\rm GeV})$ depending on the containment of the final state particles from the neutrino interaction~\cite{Super-Kamiokande:2015qek} and from the stochastic energy losses~\cite{IceCube:2013dkx}.
Though photo-coverage at the level of Super-Kamiokande is difficult to achieve, the photo-coverage of UNDINE can be chosen as a trade-off between construction costs and event reconstruction capability.

Both detectors are very cost-effective.
UNDINE can leverage lake Geneva for water overburden, and, given the CHIPS cost estimates presented in Ref.~\cite{CHIPS:2024vpb}, can likely be deployed for under US\$10M.
The surface-based detection strategy of SINE alleviates the need to excavate a cavern, as is the case for the FPF~\cite{Adhikary:2024nlv}.
Though a full cost-estimate of SINE is out of the scope of this letter, the simple detector design and lack of a cavern should keep costs below those of UNDINE.
These figures should be compared to the expected US\$100M total cost of the FPF~\cite{Adhikary:2024nlv}.

\section{Neutrino Interaction Rates} \label{sec:rates}

\begin{figure*}[t]
    \centering
    \includegraphics[width=0.9\linewidth]{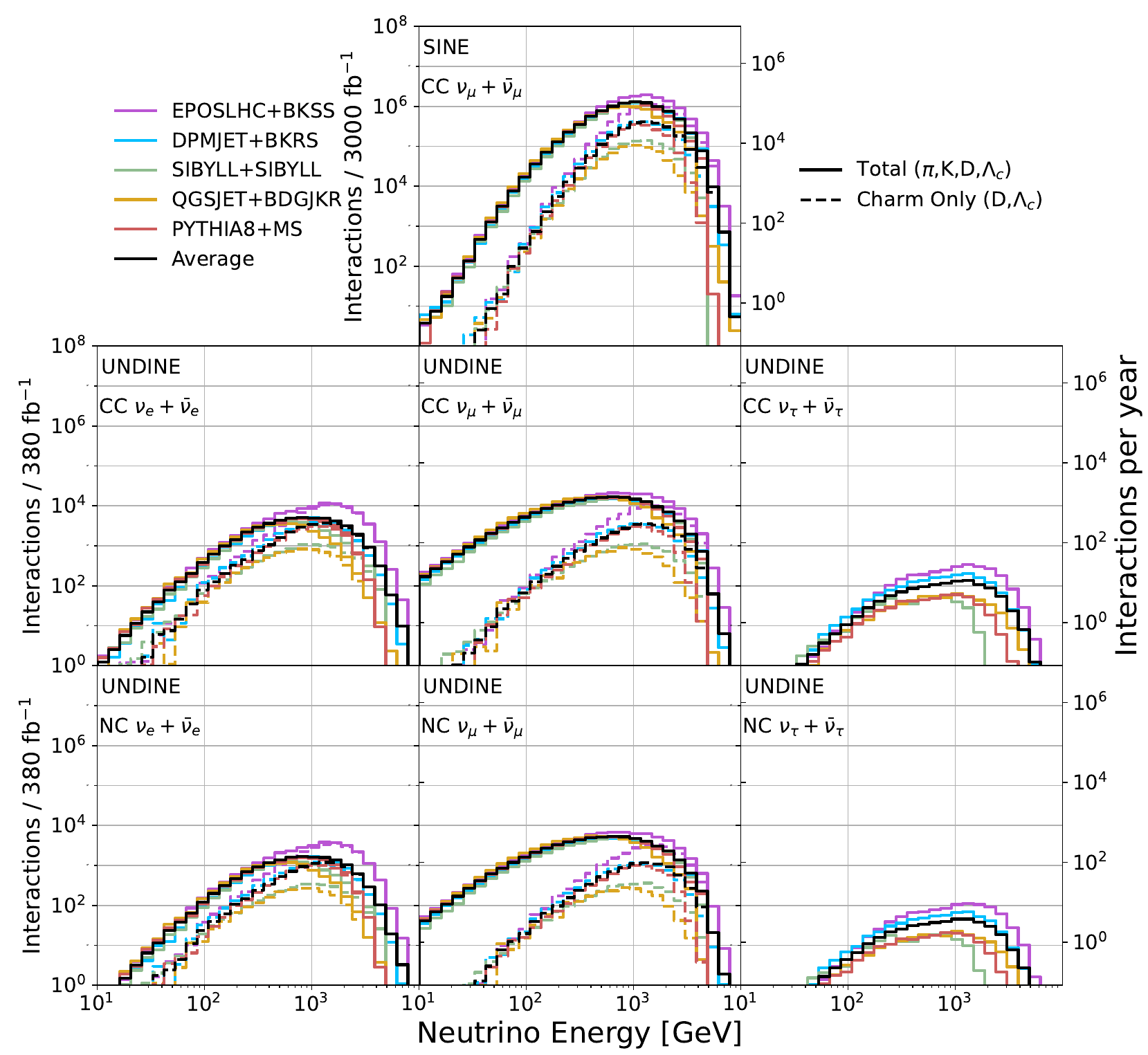}
    \caption{\textbf{\textit{Energy distribution of events.}} Differential event rates as a function of neutrino energy in SINE and UNDINE. Each column corresponds to a different neutrino flavor, while each row assumes a different detector and interaction type: CC DIS in SINE (top), CC DIS in UNDINE (middle), and NC DIS in UNDINE (bottom).
    Each sub-figure shows the energy distribution for five different hadron-production models of the forward neutrino flux. 
    The contribution from charm hadrons is shown separately.}
    \label{fig:energy-distributions}
\end{figure*}

\bgroup
\def\arraystretch{1.2}
\begin{table}[b]
    \centering
    \begin{tabular}{|c|c|c|c|}
    \hline
    Dataset & \bf{Total} & $\pi,\,K$ & $D,\,\Lambda_c$  \\
    \hline
    \hline    
SINE (CC $\nu_\mu+\bar{\nu}_\mu$) & $10^{6.98}$ & $10^{6.84}$ & $10^{6.40}$ \\
UNDINE (CC $\nu_e+\bar{\nu}_e$) & $10^{4.68}$ & $10^{4.32}$ & $10^{4.42}$ \\
UNDINE (CC $\nu_\mu+\bar{\nu}_\mu$) & $10^{5.27}$ & $10^{5.20}$ & $10^{4.41}$ \\
UNDINE (CC $\nu_\tau+\bar{\nu}_\tau$) & $10^{3.07}$ & 0 & $10^{3.07}$ \\
UNDINE (NC $\nu_\alpha + \bar{\nu}_\alpha$) & $10^{4.87}$ & $10^{4.76}$ & $10^{4.24}$ \\
    \hline
\end{tabular}
    \caption{\textbf{\textit{SINE and UNDINE event rates.}} Expected number of recorded neutrino interactions for SINE and UNDINE in $3000\,{\rm fb}^{-1}$ and $380\,{\rm fb}^{-1}$, respectively, corresponding to the expected integrated luminosity of HL-LHC. Event rates are averaged over the hadron-production models described in \cref{sec:rates}.}
\label{tab:event_rates}
\end{table}
\egroup

We use the \SIREN~simulation package to calculate the number of neutrino interactions expected at SINE and UNDINE~\cite{Schneider:2024eej}.
We inject neutrinos according to publicly available simulated samples of the ATLAS forward neutrino flux~\cite{makelat,Kling:2023tgr}, described further in \cref{app:flux}.
These samples are separated into neutrinos produced in the decays of light mesons ($\pi$,~$K$) and charm hadrons ($D$,~$\Lambda_c$).
Multiple samples for different hadron-production models are provided in each case: EPOS-LHC~\cite{Pierog:2013ria}, DPMJET-III~\cite{Roesler:2000he,Fedynitch:2015kcn}, SIBYLL 2.3c~\cite{Fedynitch:2018cbl}, QGSJET-II~\cite{Ostapchenko:2010vb}, and the forward tune of Pythia 8.2~\cite{Fieg:2023kld} for the pion and kaon parents; and BKSS kT~\cite{Bhattacharya:2023zei}, BKRS~\cite{Buonocore:2023kna}, SIBYLL 2.3c~\cite{Fedynitch:2018cbl}, BDGJKR~\cite{Bai:2022xad}, and MS kT~\cite{Maciula:2022lzk} for the charm hadron parents.
We compute the event rate separately for each hadron-production model.
Hereafter, we compute total event rates by matching each light meson sample with a specific charm hadron sample following the convention in Ref.~\cite{Kling:2023tgr}.
Each sample contains a list of neutrinos and antineutrinos with their flavor, four momentum, production location, parent hadron, and physical weight.
\SIREN~then simulates the deep-inelastic scattering (DIS) of these neutrinos along the forward neutrino beamline.
We use the total and differential neutrino DIS cross sections computed in Ref.~\cite{Weigel:2024gzh}.

For SINE, we consider $\nu_\mu$ and $\overline{\nu}_\mu$ CC DIS in the bedrock up to approximately $5\,{\rm km}$ upstream of the scintillator detector.
To determine whether the final state muon reaches the detector, we use \SIREN~to reject muons that do not intersect the scintillator panels or for which the traversed column depth of the muon is greater than the muon range calculated in Ref.~\cite{Chirkin:2004hz}.
For UNDINE, we consider all-flavor CC and neutral-current (NC) DIS within the detectors themselves.
For the moment, we ignore the additional contribution of through-going muons from $\nu_\mu$ and $\overline{\nu}_\mu$ interactions outside of UNDINE.
We assume 100\% detection efficiency for all muons the pass through SINE and all neutrinos that interact in UNDINE.
This is a reasonable approximation, as scintillation-based~\cite{MINERvA:2013zvz} and Cherenkov-based~\cite{IceCube:2011ucd} detectors have demonstrated detection efficiencies of approximately 100\% for tagging neutrino interactions at energies similar to or below those expected at SINE and UNDINE.
We also neglect uncertainties on this efficiency for this study, which will be subdominant to flux and cross section uncertainties as long as they can be kept at or below the percent level.


In \cref{tab:event_rates} we report the expect neutrino event rates in SINE and UNDINE over the course of HL-LHC.
We consider an expected integrated luminosity of $3000\,{\rm fb}^{-1}$ for CMS/SINE and $380\,{\rm fb}^{-1}$ for LHCb/UNDINE~\cite{Drewes:2019fou}.
We expect $\mathcal{O}(10{\rm M})$ and $\mathcal{O}(0.1{\rm M})$ total neutrino interactions recorded in SINE and UNDINE, respectively.
\Cref{fig:energy-distributions} shows the energy distribution of these neutrino interactions in each detector.
We show separately the rate from charm-produced neutrinos, which dominate the total neutrino interaction rate at higher energies and exhibit large differences between the five hadron-production models listed.
One can also see that SINE has a harder neutrino spectrum than UNDINE; this is because of the added impact of the muon range.
Thus, charm-produced neutrinos comprise a larger fraction of the SINE neutrino dataset than the UNDINE neutrino dataset.

\section{Cosmic Muon Background Rates and Mitigation}
\label{sec:bkg}

\begin{figure}[ht]
    \centering
    \includegraphics[width=0.8\linewidth]{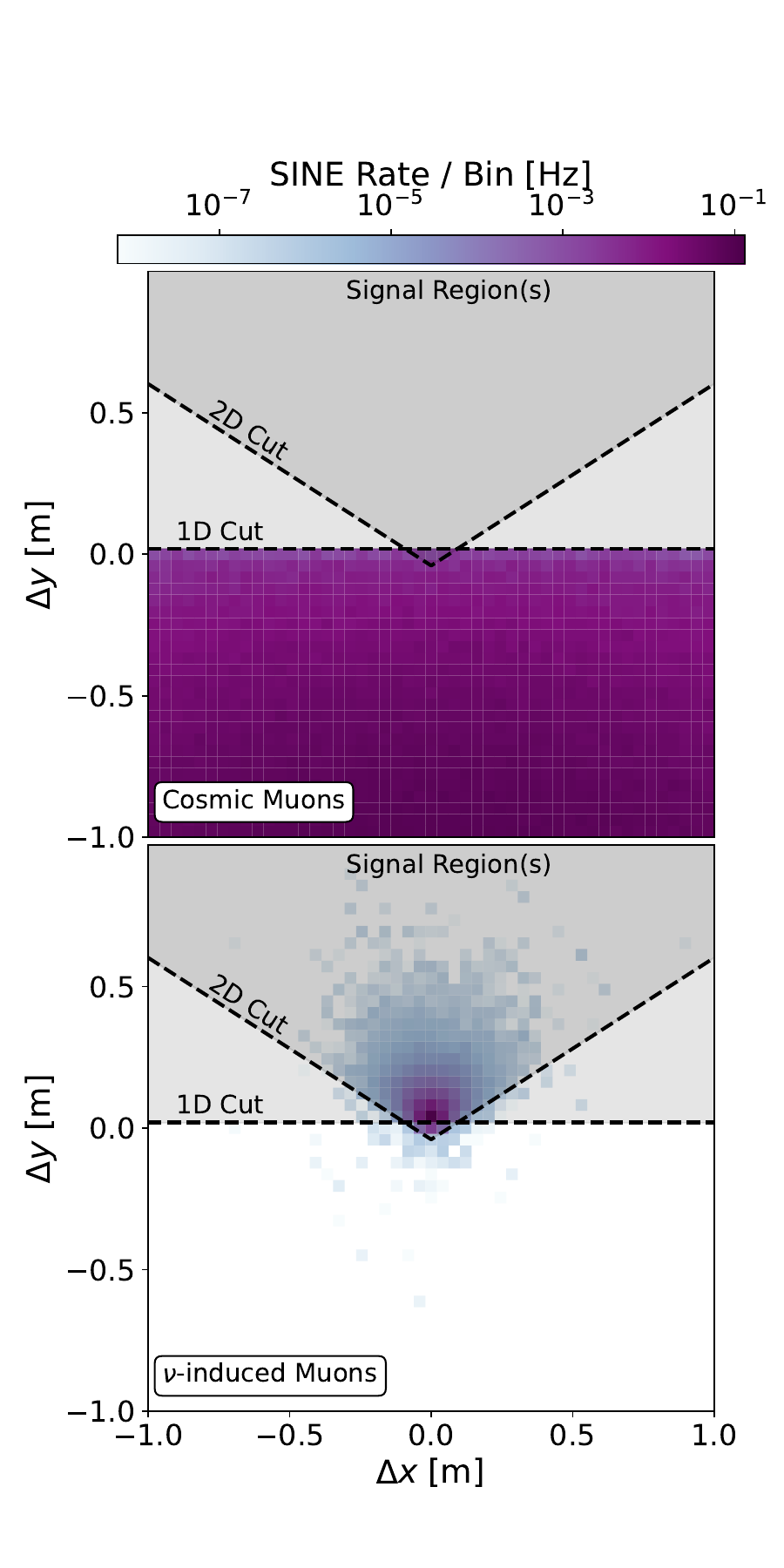}
    \caption{\textbf{Transverse displacement of signal and background in SINE.} The distribution of $(\Delta x,\Delta y)$, the horizontal and vertical displacement between the front and back panels in SINE, for cosmic muon backgrounds (top) and neutrino-induced muon signals (bottom). The color axis indicates the rate during HL-LHC operation. The dotted lines and shaded regions indicate one and two-dimensional spatial cuts to separate signal from background.}
    \label{fig:spatial_distributions}
\end{figure}

We now discuss cosmic ray muon backgrounds to neutrino interactions at SINE and UNDINE.
In the latter case, the CHIPS collaboration has demonstrated that that 50\,m of water overburden is sufficient to separate neutrino interactions from cosmic muon backgrounds in a water Cherenkov detector~\cite{Tingey:2022evd}.
Given that the neutrino event rates in UNDINE are higher compared to CHIPS, it is likely that similar techniques can be employed to entirely reject cosmic muon backgrounds at UNDINE.

In SINE, we consider a series of timing and spatial cuts to separate cosmic muons from neutrino-induced muons.
We use \textsc{EcoMug}~\cite{Pagano:2021wqu} to simulate cosmic-ray muon backgrounds in one SINE sub-detector, i.e. one stack of $2\times3$ shipping containers.
Before any cuts, cosmic muons intersect both scintillator panels at a rate of approximately 5\,kHz.
We rely on two timing cuts and two spatial cuts to reduce this background rate to below the rate of neutrino-induced muons in SINE during HL-LHC operation, approximately 0.1\,Hz.
The timing cuts are possible thanks to the $\mathcal{O}$(ns) timing resolution of plastic scintillator, while the spatial cuts are possible with $\mathcal{O}$(cm) segmentation of the scintillator panels in one or two dimensions.

The first timing leverages the timing structure of $pp$ collisions at the LHC, which are spaced by 25\,{\rm ns}~\cite{Taylor:592719}.
Over 99\% of the neutrino-induced muons arrive within 2.5\,ns of the collision.
Defining this as our signal window and accounting for missing bunches~\cite{Taylor:592719}, this leads to a duty factor of 7.9\%, reducing the cosmic background rate to 395\,Hz.
The second timing cut leverages the time between muon intersections with the front and back scintillator panels.
As neutrino-induced muons tend to travel approximately perpendicular through the panels, they always result in a panel time difference of $8 < \Delta t/{\rm ns}<9$.
In contrast, cosmic muons are downward-going in zenith and uniformly distributed in azimuth, resulting in longer typical panel time differences or a reversed panel crossings order, i.e. $\Delta t < 0$.
We define our signal region to be $8 < \Delta t/{\rm ns}<10$, reducing the cosmic background rate to 135\,Hz.

We next consider spatial cuts to better leverage the nearly-perpendicular path of neutrino-induced muons.
\Cref{fig:spatial_distributions} shows the distribution of transverse displacement between the front and back panels, $(\Delta x, \Delta y)$, for neutrino-induced muons and cosmic muons.
The former are slightly up-going, while the latter are predominately down-going.
We thus consider the one and two-dimensional cuts displayed in \cref{fig:spatial_distributions}.
The former, defined by the region $\Delta y>2\,{\rm cm}$, rejects all simulated cosmic muons while retaining 98.1\% of neutrino-induced muons.
The latter, defined by the triangular region in \cref{fig:spatial_distributions}, rejects most cosmic muons while retaining 99.3\% of neutrino-induced muons, resulting in a background rate of 2\,mHz.

\begin{figure}[ht]
    \centering
    \includegraphics[width=\linewidth]{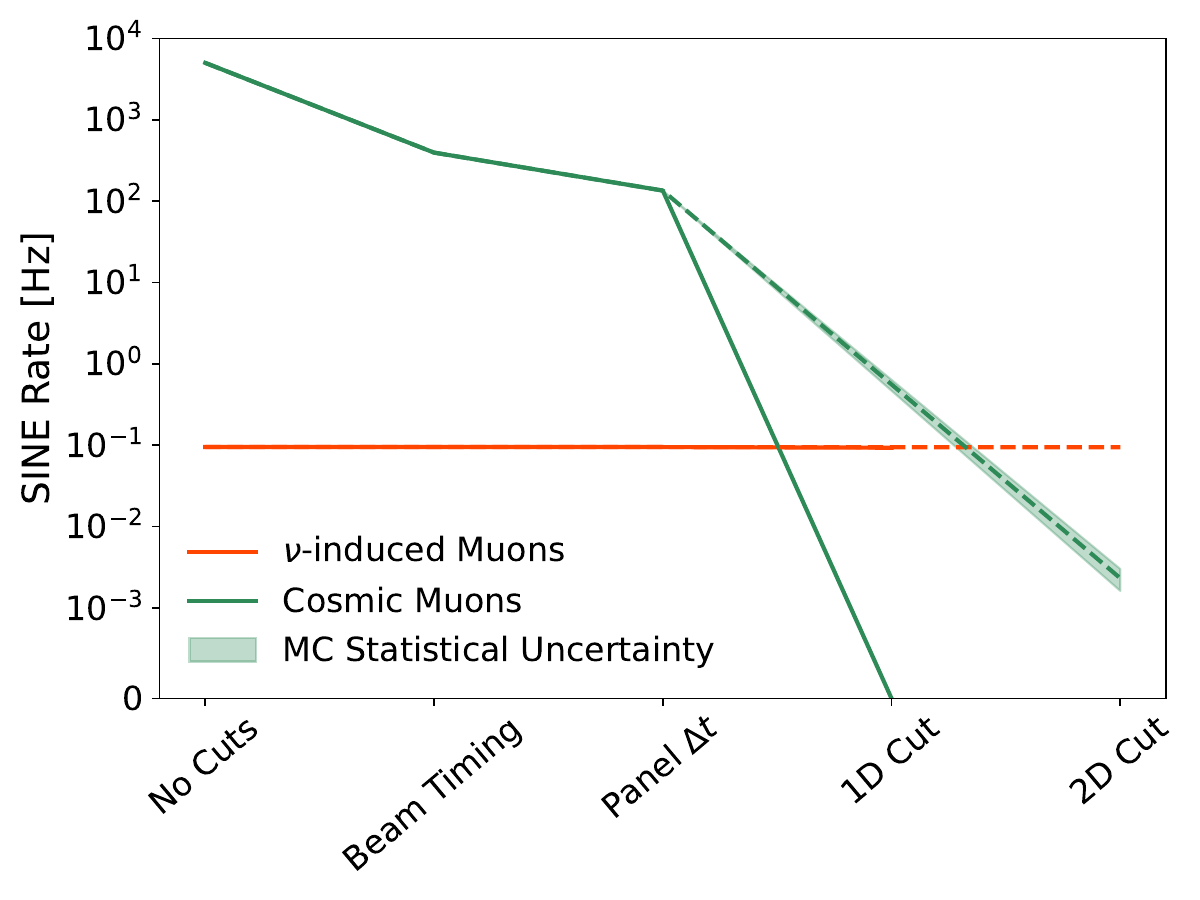}
    \caption{\textbf{Cosmic muon mitigation in SINE.} The neutrino-induced muon signal rate and the cosmic muon background rate over the series of timing and spatial cuts described in \cref{sec:bkg}. The solid (dashed) line corresponds to the one (two) dimensional spatial cut. The shaded band around the cosmic background rate indicates the statistical uncertainty from the \textsc{EcoMug} simulation.}
    \label{fig:SINE_backgrounds}
\end{figure}

\Cref{fig:SINE_backgrounds} summarizes the signal and background rates in SINE following the cuts described above.
The two timing cuts in combination with either spatial cut will be sufficient to achieve a negligible cosmic muon background rate in SINE.

\section{Physics Opportunities} \label{sec:physics}

The large datasets collected by SINE and UNDINE provide fertile ground to study various physics scenarios.
We focus here on four first studies: neutrino DIS cross sections, charm-hadron production in $pp$ collisions, strangeness enhancement in $pp$ collisions, and sensitivity to heavy neutral leptons.
Future work will assess the sensitivity to other long-lived particles targeted by FASER and the proposed FPF experiments~\cite{Anchordoqui:2021ghd}, as well as rare neutrino-nucleus interactions leading to di-muons in SINE.
These can arise from the production of charm hadrons in neutrino DIS, allowing for a unique measurement of the strange parton distribution function (PDF)~\cite{Cruz-Martinez:2023sdv}.

\subsection{Neutrino Cross Sections}

\begin{figure}[ht]
    \centering
    \includegraphics[width=1.\linewidth]{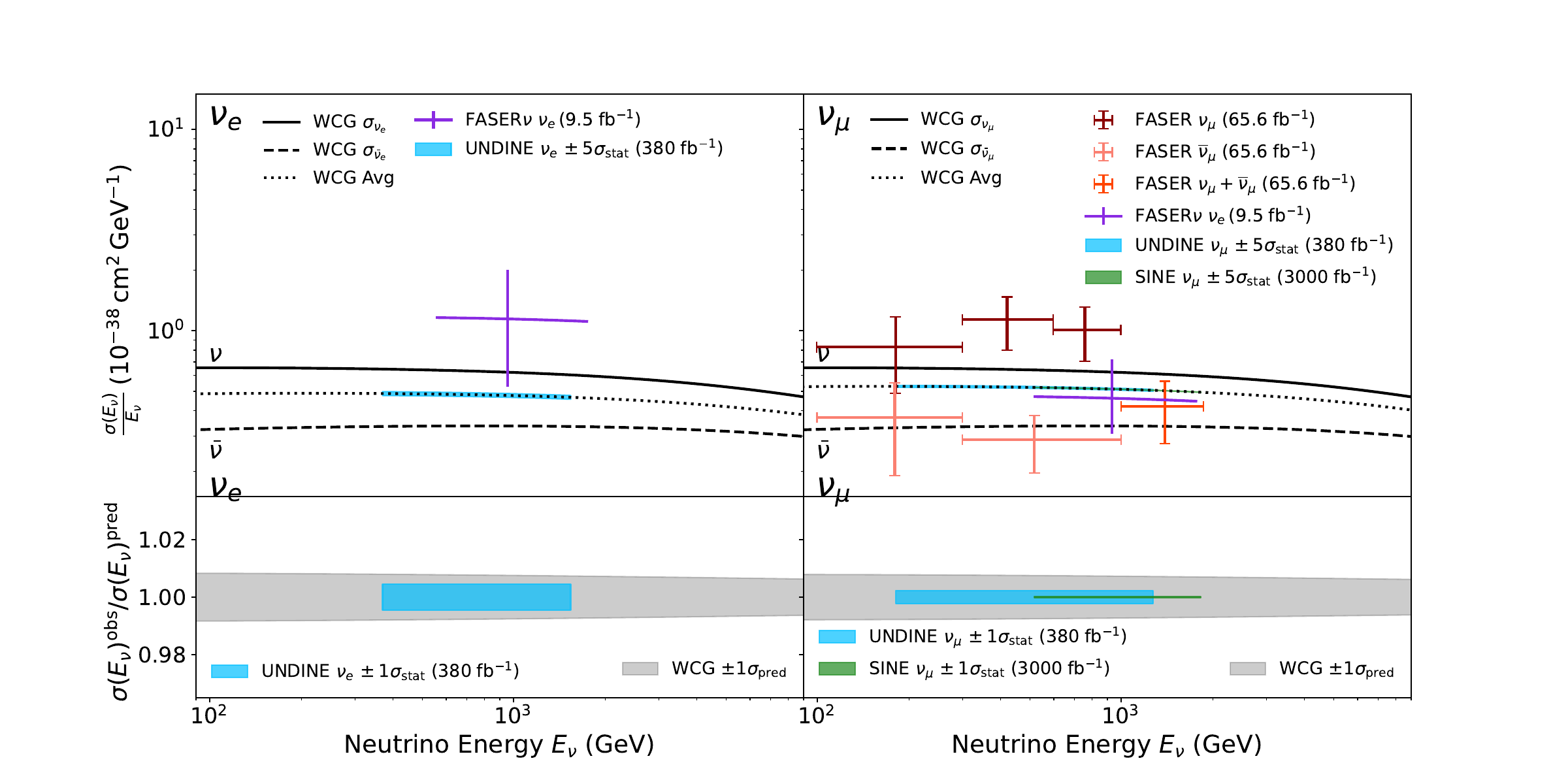}
    \caption{\textbf{\textit{Cross section sensitivity.}} Expected statistical uncertainties on a one-bin total $\nu_e$ (left) and $\nu_\mu$ (right) cross section measurement with the full HL-LHC dataset. SINE can measure the $\nu_\mu$ cross section, while UNDINE can independently measure the $\nu_\mu$ and $\nu_e$ cross sections. The top panels show expected uncertainties on the average $\nu + \overline{\nu}$ cross section for each flavor with the full HL-LHC dataset. Recent measurements from FASER are also shown~\cite{FASER:2024ref,FASER:2024hoe}. The bottom panels show the relative measurement uncertainties compared to the PDF uncertainties on the DIS cross section in Ref.~\cite{Weigel:2024gzh}, represented here by the ``WCG'' band.}
    \label{fig:cross-section}
\end{figure}

Neutrinos from the LHC offer a unique opportunity to measure neutrino DIS cross sections at TeV energies.
The large datasets of SINE and UNDINE allow for competitive measurements of these cross sections.
Both detectors can measure the $\nu_\mu + \overline{\nu}_\mu$ CC DIS cross section.
UNDINE will also be able to measure $\nu_e + \overline{\nu}_e$ CC DIS cross section due to the particle identification capability of water Cherenkov detectors.
\Cref{fig:cross-section} shows the cross section sensitivity of SINE and UNDINE after one year of HL-LHC data.
We present a one-bin measurement of the total neutrino DIS cross section, though in principle UNDINE will be able to resolve the initial neutrino energy~\cite{Super-Kamiokande:2015qek}.
Statistical uncertainties are at the $\mathcal{O}(1\%)$ level, comparable to the PDF uncertainties of Ref.~\cite{Weigel:2024gzh}.
Studies of individual hadron-production models suggest flux uncertainties of approximately $20$-$40$\% on LHC neutrino cross section measurements~\cite{Fieg:2023kld,Buonocore:2023kna}.
These flux uncertainties dominate over statistical uncertainties as well as other sources of systematic uncertainty~\cite{FASER:2024ref}, and would be similar between our detectors and other proposed detectors, including FASER$\nu$2 and FLOUNDER~\cite{Ariga:2025jgv}.
Thus, SINE and UNDINE can perform competitive measurements of total neutrino cross sections.

\subsection{Constraining Charm Production in $pp$ Collisions}

Due to the limited pseudorapidity coverage of LHC detectors, forward neutrino experiments are uniquely sensitive to hadron production in $pp$ collisions in the forward direction ($\eta \gtrsim 9$)~\cite{FASER:2022hcn}.
The lack of prior measurements leads to large discrepancies between predictions of the forward neutrino flux from different hadron-production models, especially for charm-flavored hadrons, as shown in \cref{fig:energy-distributions}.
SINE and UNDINE offer a novel avenue to distinguish between these models through two ratio measurements: the total rate in SINE v.s. UNDINE and the rate of $\nu_\mu$ v.s. $\nu_e$ interactions in UNDINE.
Both are sensitive to charm hadron contribution to the neutrino flux; the former because charm hadrons comprise a larger fraction of the total neutrino interaction rate in SINE compared to UNDINE, and the latter because charm hadrons comprise a larger fraction of the $\nu_e$ flux compared to the $\nu_\mu$ flux.
We emphasize that the first ratio is a completely independent method through which to probe charm production compared to the FPF, enabled by SINE's unique ability to tag neutrino interactions with through-going muons.

\begin{figure}[ht]
    \centering
    \includegraphics[width=\linewidth]{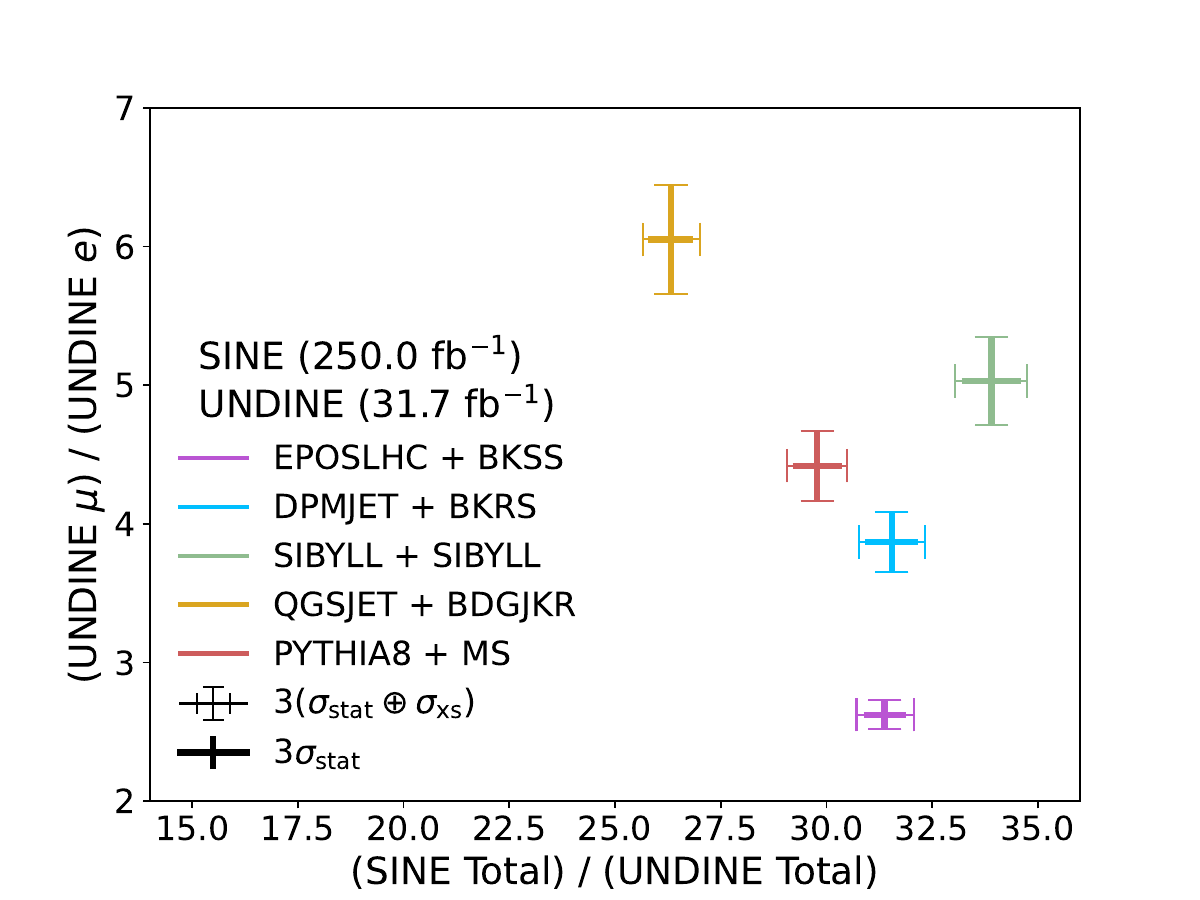}
    \caption{\textbf{\textit{Ratio measurements to distinguish hadron-production models.}} Ratios of the $\nu_\mu$ to $\nu_e$ interaction rate in UNDINE and the total event rate in SINE to UNDINE, as predicted by the different hadron-production models. The error bars represent the expected statistical and cross section $3\sigma$ uncertainties after one year of HL-LHC data.}
    \label{fig:generator-differences}
\end{figure}

\Cref{fig:generator-differences} shows the ratios predicted by each of the hadron-production models.
The error bars here represent the expected statistical and cross section uncertainties under each model hypothesis, where the latter come from Ref.~\cite{Weigel:2024gzh}.
After one year of the HL-LHC, SINE and UNDINE data can distinguish between the hadron-production models at the $\gtrsim 3\sigma$ confidence level.
This especially interesting given the recent measurement of the muon neutrino flux at FASER, which exhibits around $2\sigma$ tension with all hadron-production models~\cite{FASER:2024ref}.
Complementary measurements of the LHC forward neutrino flux from SINE and UNDINE will be essential if this tension persists. 

\subsection{Strangeness Enhancement} \label{sec:strangeness}

SINE and UNDINE data are also sensitive to enhanced strange meson production in the forward direction of $pp$ collisions.
Such an enhancement has been proposed as a solution to the cosmic-ray muon puzzle---an approximately $8\sigma$ excess of muons observed in cosmic-ray air showers with respect to predictions from hadronic shower models~\cite{Albrecht:2021cxw,Anchordoqui:2022fpn}.
This scenario would increase the forward pion-to-kaon ratio in $pp$ collisions and would thus have an observable effect on forward neutrino fluxes~\cite{Albrecht:2021cxw,Anchordoqui:2022fpn,Kling:2023tgr}.

\begin{figure}[ht]
    \centering
    \includegraphics[width=\linewidth]{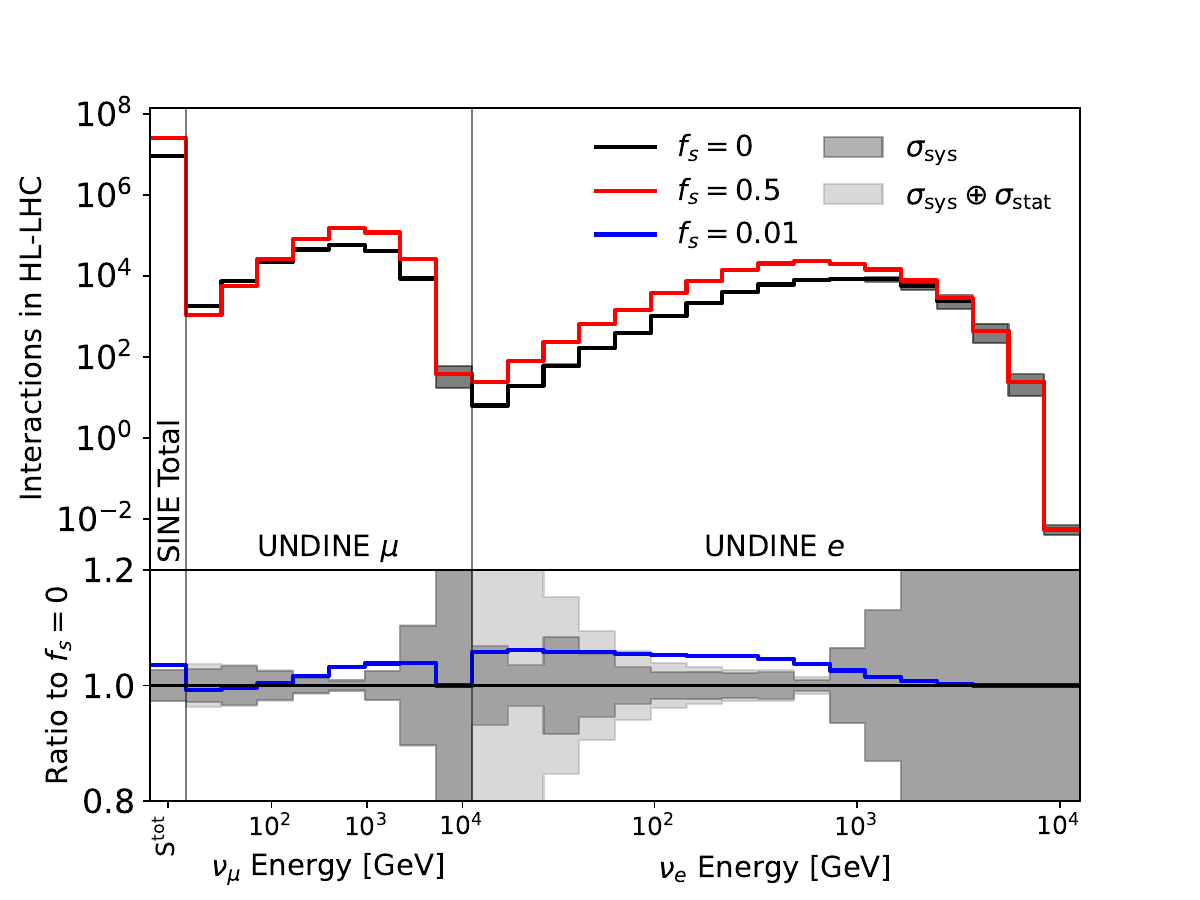}
    \includegraphics[width=\linewidth]{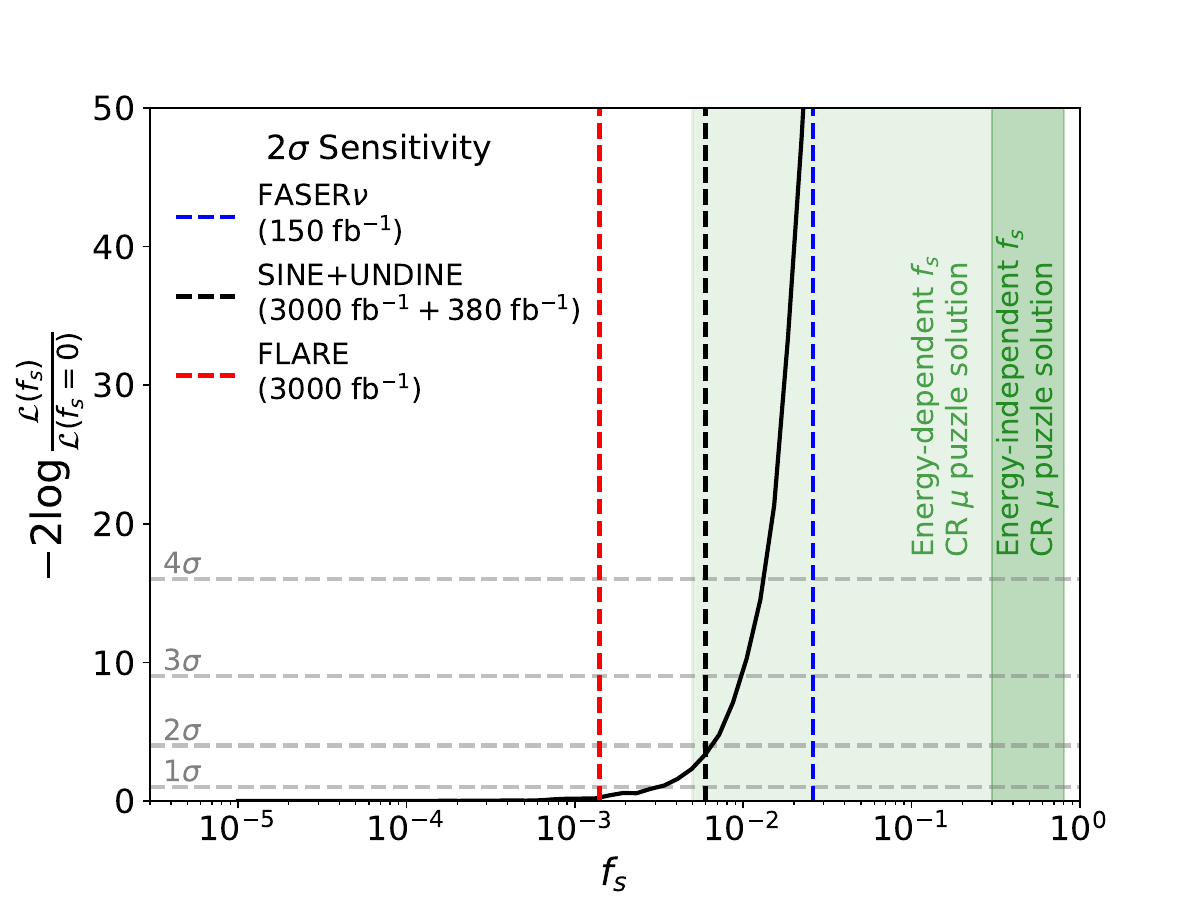}
    \caption{\textbf{\textit{Strangeness enhancement sensitivity.}} The top panel shows the expected SINE total event rate and the UNDINE differential event rate in neutrino energy separately for $\nu_\mu$ and $\nu_e$ interactions for $f_s=0$, $f_s=0.5$, and $f_s=0.01$. Note that the $f_s=0.01$ curve lies behind the $f_s = 0$ curve in the interaction rate subplot, while the $f_s=0.5$ curve lies outside of the displayed range in the ratio subplot. Also shown are the expected statistical and flux systematic uncertainties in each bin, where the latter are derived using the procedure discussed in Ref.~\cite{Kling:2023tgr}. The bottom panel shows the log-likelihood ratio as a function of $f_s$ for a SINE and UNDINE analysis. The expected $2\sigma$ sensitivity is shown by the vertical black line. Also shown are the expected FASER$\nu$ and FLArE sensitivity after the full LHC Run 3 and HL-LHC, respectively. The green regions represent the energy-independent and energy-dependent solutions to the cosmic muon puzzle~\cite{Anchordoqui:2022fpn,Kling:2023tgr,Sciutto:2023zuz}.} 
    \label{fig:strangeness-variation}
\end{figure}

Following the procedure outlined in Ref.~\cite{Kling:2023tgr}, we introduce a pion-to-kaon swapping probability $f_s$, such that the number of pion-produced neutrinos is reduced by $N_{\pi \to \nu} \to (1 - f_s) N_{\pi \to \nu}$ and the number of kaon-produced neutrinos is enhanced by $N_{K \to \nu} \to (1 + 6.6 f_s) N_{K \to \nu}$.
We then compute the number of events in SINE and UNDINE as a function of $f_s$.
We consider only the total rate in SINE, while for UNDINE we consider separate muon and electron samples binned according to the initial neutrino energy~\footnote{We use 8 and 17 bins across three decades in neutrino energy for the UNDINE muon and electron samples, respectively. The difference is because the electron sample is expected to have better neutrino energy resolution.}.
The top panel of \cref{fig:strangeness-variation} shows the expected number of events in these datasets for $f_s=0$, $f_s=0.5$, and $f_s=0.01$.
For illustration purposes, we also show the expected statistical and systematic flux uncertainties.
The latter are computed following the procedure outlined in Appendix A of Ref.~\cite{Kling:2023tgr}: interpolation parameters are introduced between the different hadron-production models in \cref{sec:rates}, and the per-bin uncertainty is computed by the Cram{\'e}r-Rao bound, in which the covariance matrix is the inverse of Fisher information matrix comprising the interpolation parameters and the rate in each bin of the top panel of \cref{fig:strangeness-variation}.
One can see that a non-zero $f_s$ can alter the UNDINE $\nu_\mu$ and $\nu_e$ energy distributions in a way that is not captured by the flux systematic uncertainties.

In the bottom panel of \cref{fig:strangeness-variation}, we show the Asimov sensitivity of SINE and UNDINE to the $f_s$ parameter.
This is computed using a log-likelihood ratio test statistic considering a binned Poisson likelihood over the distributions in the top panel of \cref{fig:strangeness-variation}.
To account for flux uncertainties, we profile over the hadron-production model interpolation parameters introduced in Ref.~\cite{Kling:2023tgr}.
This analysis suggests that SINE and UNDINE will be able to constrain $f_s \simeq 0.006$ at the $2\sigma$ confidence level.
This is much smaller than the expected range to explain the cosmic-ray muon puzzle, $f_s \approx 0.3-0.8$~\cite{Anchordoqui:2022fpn,Kling:2023tgr}.
It will also be able to rule out a scenario in which $f_s$ is energy dependent, which suggests $f_s \approx 0.005$ at LHC energies~\cite{Kling:2023tgr,Sciutto:2023zuz}.
We also show the expected $f_s$ sensitivity of FASER$\nu$ after Run 3 of the LHC and FLArE after the HL-LHC in \cref{fig:strangeness-variation}~\cite{Kling:2023tgr}.
Though SINE and UNDINE will not be as sensitive as FLArE, they will be able to perform a competitive search for strangeness enhancement that covers the relevant $f_s$ parameter space for the cosmic-ray muon puzzle.

\subsection{Heavy Neutral Lepton Searches}

Heavy neutral leptons (HNLs) are hypothetical right-handed counterparts to the left-handed neutrinos that appear commonly in theories of neutrino mass~\cite{Minkowski:1977sc,Schechter:1980gr}, leptogenesis~\cite{Fukugita:1986hr}, and dark matter~\cite{Dodelson:1993je}; see Ref.~\cite{Abdullahi:2022jlv} for a recent review.
SINE can search for HNLs produced through two different channels: meson decays near the interaction point, and neutrino upscattering in the bedrock between the interaction point and the detector.
We use the \textsc{FORESEE} package to simulate HNL production from the decay of light, charm, and B mesons~\cite{Kling:2021fwx}, and \textsc{SIREN}~\cite{Schneider:2024eej} to simulate HNL production from neutrino upscattering, considering a benchmark neutrino flux predicted by the Pythia 8.2~\cite{Fieg:2023kld} and MS kT~\cite{Maciula:2022lzk} hadron-production models.
In both cases, we use \textsc{SIREN} to simulate HNL decays, following the calculations in Ref.~\cite{Ballett:2019bgd} to compute branching ratios and simulate final state kinematics.
On the detection end, we consider three strategies to separate muons from HNL decays in SINE from neutrino-induced muons, two of which were introduced in Ref.~\cite{tevpa,fpc}.
These include: (1) requiring a time delay $\Delta t >1\,{\rm ns}$ with respect to the $pp$ collision, since HNL-induced muons arrive later than neutrino-induced muons, (2) tagging HNL decays to di-muons~\cite{Chang:1975rh}, and (3) requiring a transverse displacement between panel hits $\Delta x_T > 0.1\,{\rm m}$ to isolate higher-angle HNL-induced muons.

\begin{figure}[h]
    \centering
    \includegraphics[width=\linewidth]{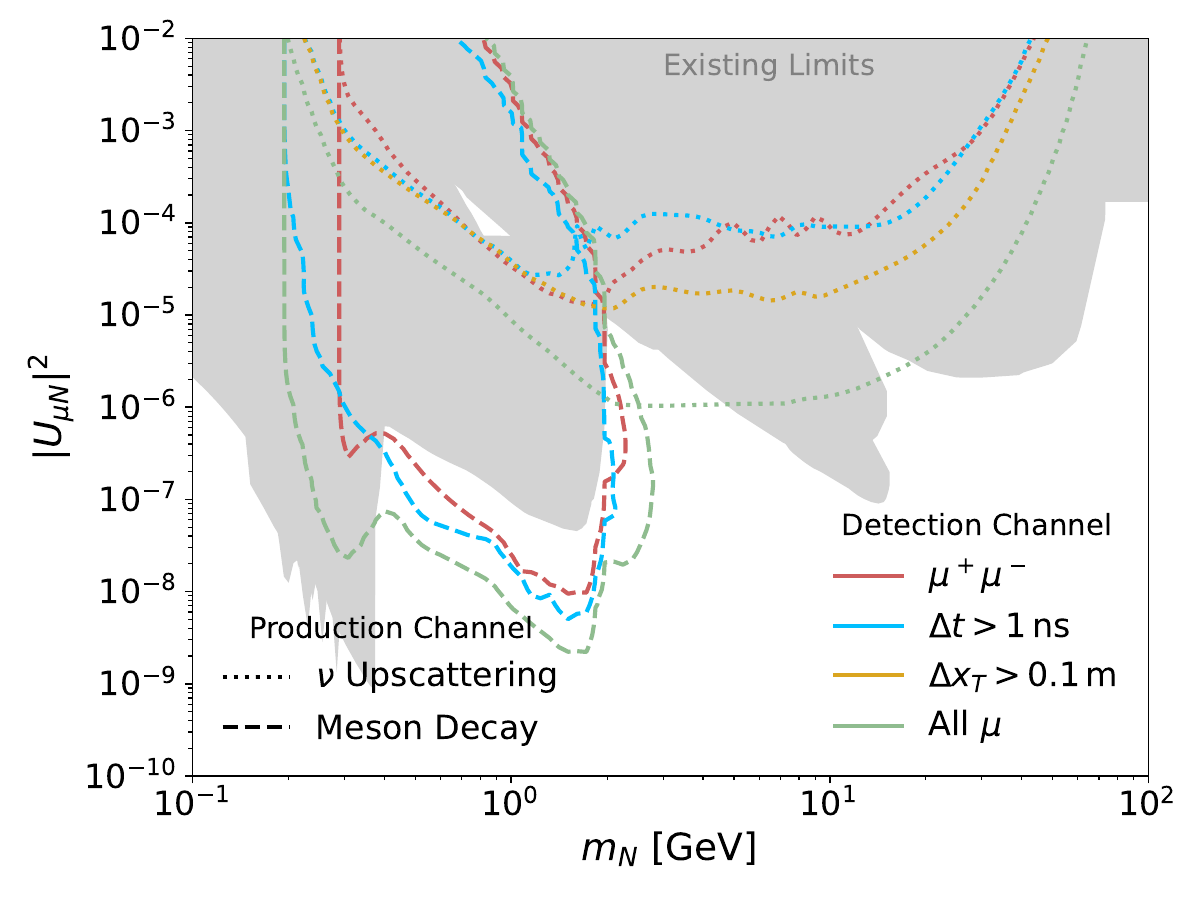}
    \caption{\textbf{Heavy neutral lepton sensitivity.} The contours indicate the HNL masses and mixings that lead to one detected muonic HNL decay in SINE throughout the HL-LHC. The different line styles indicate production via meson decays or via neutrino upscattering. The different colors indicate detection channels: dimuon decays, delayed muons $(\Delta t > 1\,{\rm ns})$, transversely-displaced muons $(\Delta x_T > 0.1\,{\rm m})$, and all HNL-induced muons. Existing limits are shown in gray and come from Ref.~\cite{Fernandez-Martinez:2023phj}}
    \label{fig:hnl}
\end{figure}

In \cref{fig:hnl}, we show contours of the HNL mass $m_N$ and muon-flavor mixing $|U_{\mu N}|^2$ that result in one HNL-induced muon in SINE for different combinations of production mechanism and detection strategy.
SINE has the potential to probe new regions of the parameter space for $m_N \lesssim 2\,{\rm GeV}$ through meson decay production and $m_N \approx 20\,{\rm GeV}$ through upscattering production, though the latter may require additional strategies to separate HNL muons from neutrino-induced muons.
It is worth emphasizing that the HNL sensitivity of SINE is competitive across nearly three orders of magnitude in $m_N$.
Such a broad HNL sensitivity is enabled by the unique design of SINE, which results in an effective decay volume of up to approximately one kilometer in front of the detector, limited by the range of muons in bedrock.
This leads to a high enough rate to probe HNLs from both meson decay and neutrino upscattering, the latter of which is not possible at the FPF due to muon backgrounds from the interaction point.
The ``broadband'' approach at SINE complements other proposed HL-LHC searches for HNLs produced in meson decays in SHiP~\cite{SHiP:2018xqw} and the FPF~\cite{Kling:2018wct}, as well as HNLs produced in the $pp$ collision itself at ATLAS, CMS and LHCb~\cite{Drewes:2019fou}, all of which probe narrower regions in $m_N$.

\section{Conclusion} \label{sec:conclusion}

In this letter, we have introduced SINE and UNDINE---two novel concepts for collider neutrino experiments at the HL-LHC.
SINE will deploy scintillator panel detectors along the CMS forward neutrino beamline to look for upward-going muons coming from $\nu_\mu$ interactions in the upstream bedrock.
UNDINE will deploy kiloton-scale water Chernekov detectors within lake Geneva along the LHCb forward neutrino beamline to look for collider neutrino interactions of all flavors.
SINE and UNDINE will record $\mathcal{O}(10{\rm M})$ and $\mathcal{O}(0.1{\rm M})$ neutrino interactions, respectively.
UNDINE compensates for its smaller sample size with its ability to distinguish neutrino flavor and improve neutrino energy reconstruction.
The physics reach of these detectors is both complementary to and competitive with the planned FPF experiments, without the need to excavate a new cavern.
We have shown that SINE and UNDINE have competitive sensitivity to TeV neutrino cross sections and the strangeness enhancement solution to the cosmic muon puzzle.
Furthermore, ratio measurements at SINE and UNDINE provide a new observable to differentiate between hadronic models of charm production in $pp$ collisions, complementing similar analyses at the FPF.
Finally, SINE has the potential to probe HNLs with competitive sensitivity across nearly three orders of magnitude in HNL mass, an important complement to next-generation HNL searches at the HL-LHC that target narrower HNL mass regions.
SINE and UNDINE thus offer a unique opportunity to establish a medium-baseline forward neutrino program at the HL-LHC.

\textit{Note added: During the completion of this letter, Ref.~\cite{Ariga:2025jgv} appeared on the arXiv pre-print server.
The results described here were obtained independently from and without knowledge of that manuscript, and preliminary results~\cite{neutrino2024,tevpa,fpc} from this study were presented for the first time in June 2024~\cite{neutrino2024}.
There are notable differences between SINE and UNDINE and the proposed detectors of Ref.~\cite{Ariga:2025jgv}; for example, the SINE proposal to use through-going muons to tag neutrino interactions in upstream bedrock is unique to this work.
That being said, the conclusions presented in Ref.~\cite{Ariga:2025jgv} also lend support for a MBF neutrino physics program at the LHC, and emphasize that, at these distances, the required target mass of detectors needs to be on the kiloton-scale.
}

\section{Acknowledgments}
We thank Albert De Roeck and Juan Rojo for helpful discussions on our proposed detectors and Benjamin Weyer for discussions on the exact beam geometries.
We thank Felix Kling for input on the forward neutrino flux models.
We also thank Philip Weigel for providing DIS cross section predictions for this letter.
We are grateful to Jackapan Pairin for providing a graphic design of our experimental concept.
Finally, we thank William Thompson for reviewing our geometry calculations and for his everlasting enthusiasm for new detectors.
C.A.A. are supported by the Faculty of Arts and Sciences of Harvard University, the National Science Foundation (NSF), the NSF AI Institute for Artificial Intelligence and Fundamental Interactions (IAIFI), the Canadian Institute for Advanced Research (CIFAR), the David and Lucile Packard Foundation, and the Research Corporation for Science Advancement.
N.W.K. is supported by the National Science Foundation (NSF) CAREER Award 2239795 and the David and Lucile Packard Foundation.
A.K. and T.Y. are supported in part by NSF grant PHY-2209445 and by the University of Wisconsin Research Committee with funds granted by the Wisconsin Alumni Research Foundation.
This work is supported by the National Science Foundation under Cooperative Agreement PHY-2019786 (The NSF AI Institute for Artificial Intelligence and Fundamental Interactions, http://iaifi.org/).

\bibliography{geneva}


\clearpage
\newpage

\onecolumngrid
\appendix

\ifx \standalonesupplemental\undefined
\setcounter{page}{1}
\setcounter{figure}{0}
\setcounter{table}{0}
\setcounter{equation}{0}
\fi

\renewcommand{\thepage}{Supplemental Methods and Tables -- S\arabic{page}}
\renewcommand{\figurename}{SUPPL. FIG.}
\renewcommand{\tablename}{SUPPL. TABLE}

\renewcommand{\theequation}{A\arabic{equation}}

\section{Supplemental Methods and Tables}

\subsection{Neutrino Flux Profile} \label{app:flux}

For this study, the forward neutrino flux at SINE and UNDINE is approximated as the ATLAS forward neutrino flux provided in Ref.~\cite{forward-nu-flux}.
We do not expect large differences in the flux at CMS and LHCb, as local magnetic fields and absorbing materials along the forward direction are similar between the different interaction points~\cite{Kling:2021gos,Buontempo:2018gta}.

\begin{figure}[ht]
    \centering
    \includegraphics[width=0.4\linewidth]{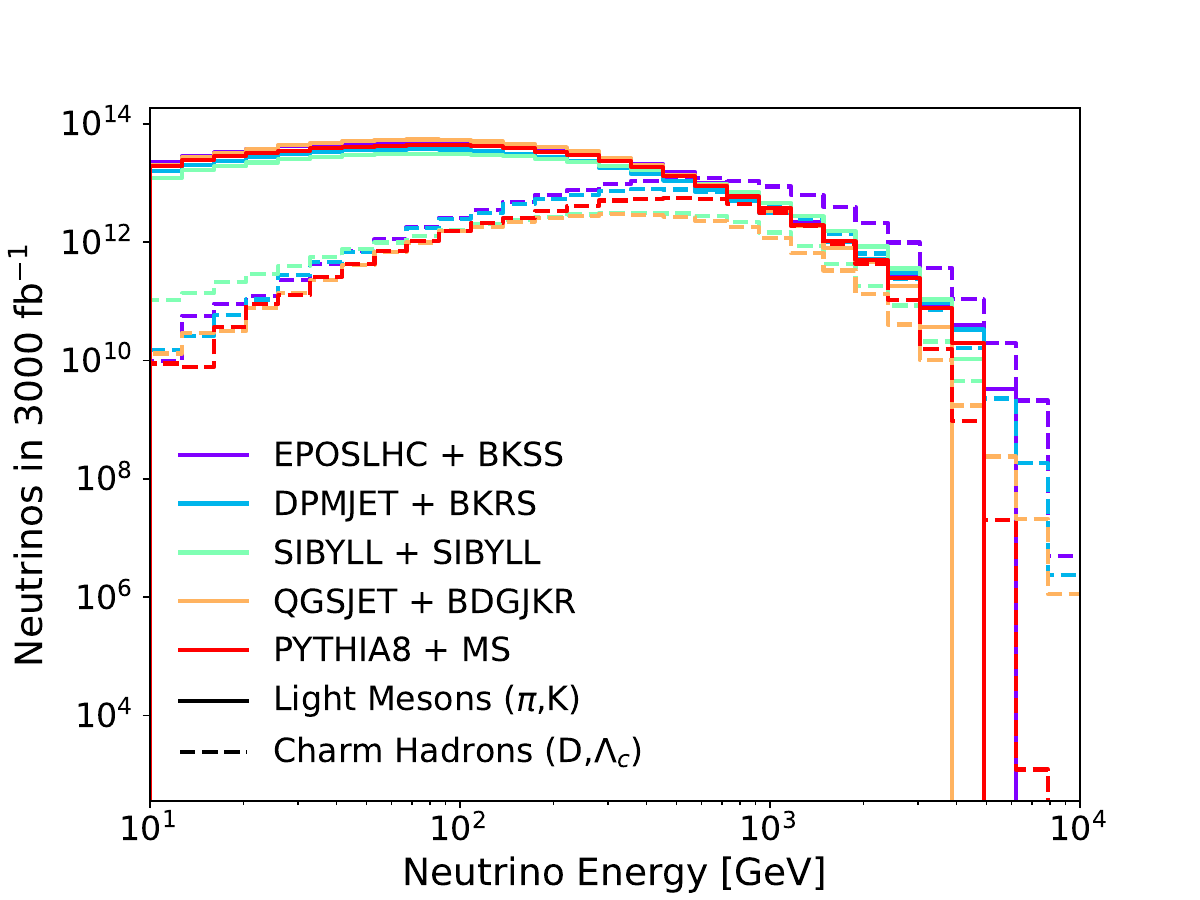}
    \includegraphics[width=0.4\linewidth]{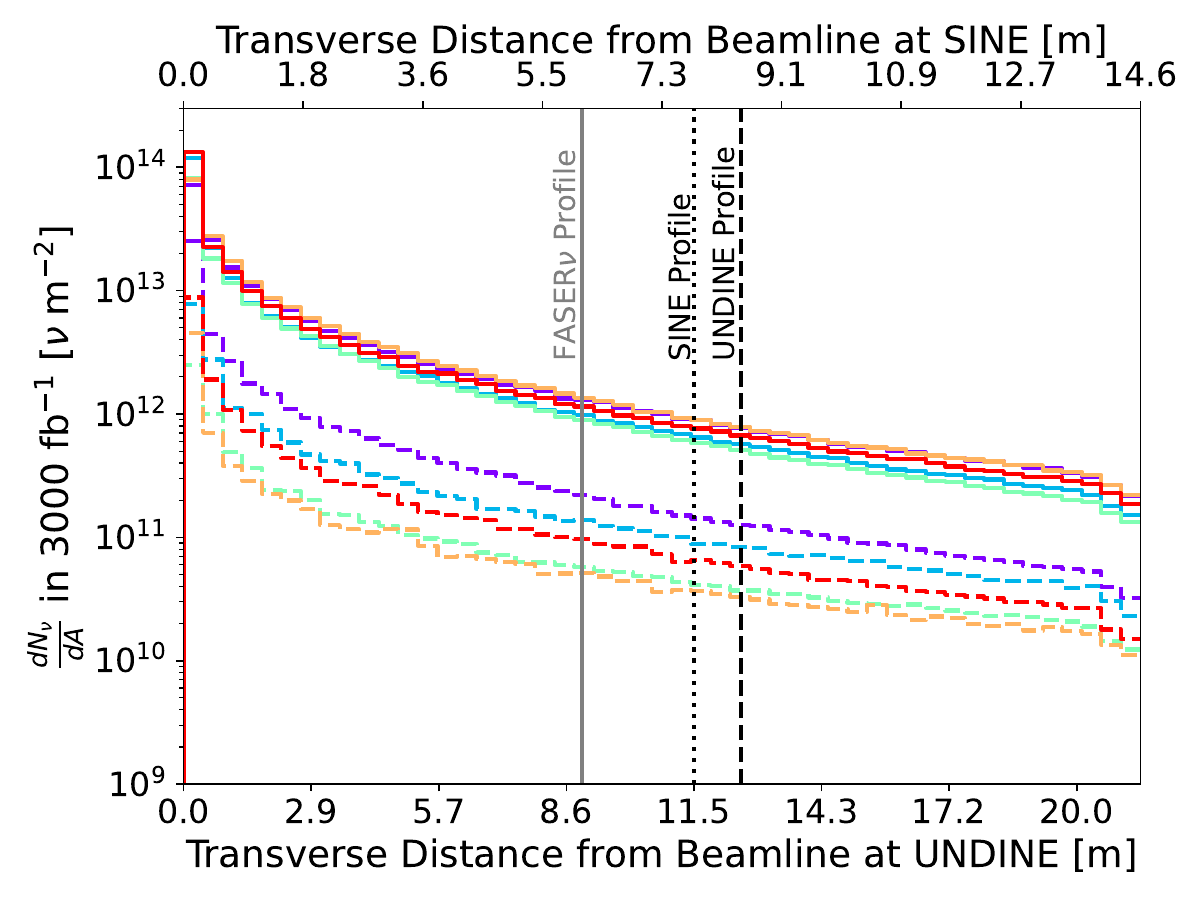}
    \caption{The energy distribution (left) and radial distribution per unit area (right) of the forward neutrino flux computed in Rev.~\cite{forward-nu-flux}. Different distributions are shown for each of the different hadron-production models introduced in \cref{sec:rates}. We also show separately neutrinos from the decay of light and charm mesons.}
    \label{fig:flux-energy-profile}
\end{figure}

The left panel of \cref{fig:flux-energy-profile} shows the energy distribution of the LHC neutrino flux for the different hadron-production models discussed in \cref{sec:rates}.
One can see order-of-magnitude uncertainties in the neutrino flux from charm hadrons, which become dominant for neutrino energies of $E_\nu \approx 1\,{\rm TeV}$.
The right panel of \cref{fig:flux-energy-profile} shows the neutrino flux per unit area as a function of transverse distance from the beamline at both SINE and UNDINE.
We also show the transverse profile of FASER$\nu$, SINE, and UNDINE, corresponding to the height of each detector: 12.5\,cm, 7.77\,m and 12.5\,m, respectively.
Despite the longer baselines of SINE and UNDINE compared to FASER$\nu$, the larger transverse profile allows each detector to envelop a larger slice of the forward neutrino flux.

\subsection{Cosmic Muon Backgrounds in SINE} \label{app:sine_backgrounds}

\begin{figure}
    \centering
    \includegraphics[width=\linewidth]{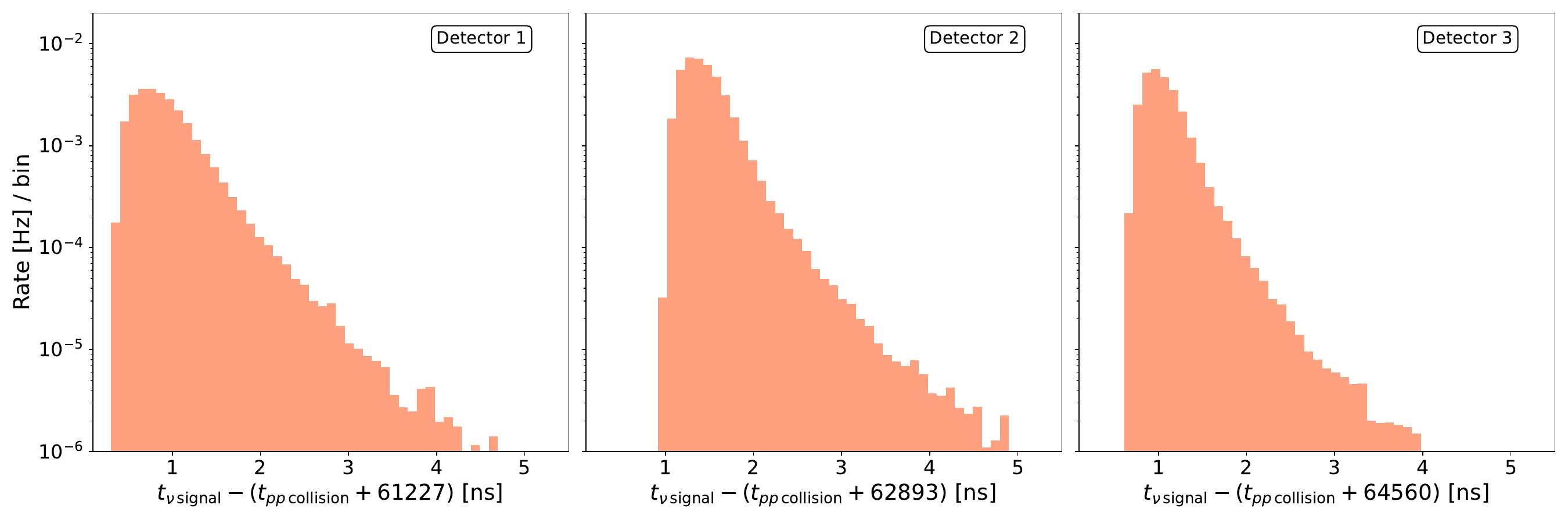}
    \caption{The distribution of neutrino-induced muon arrival times in each SINE sub-detector.}
    \label{fig:beam_timing}
\end{figure}

\begin{figure}
    \centering
    \includegraphics[width=\linewidth]{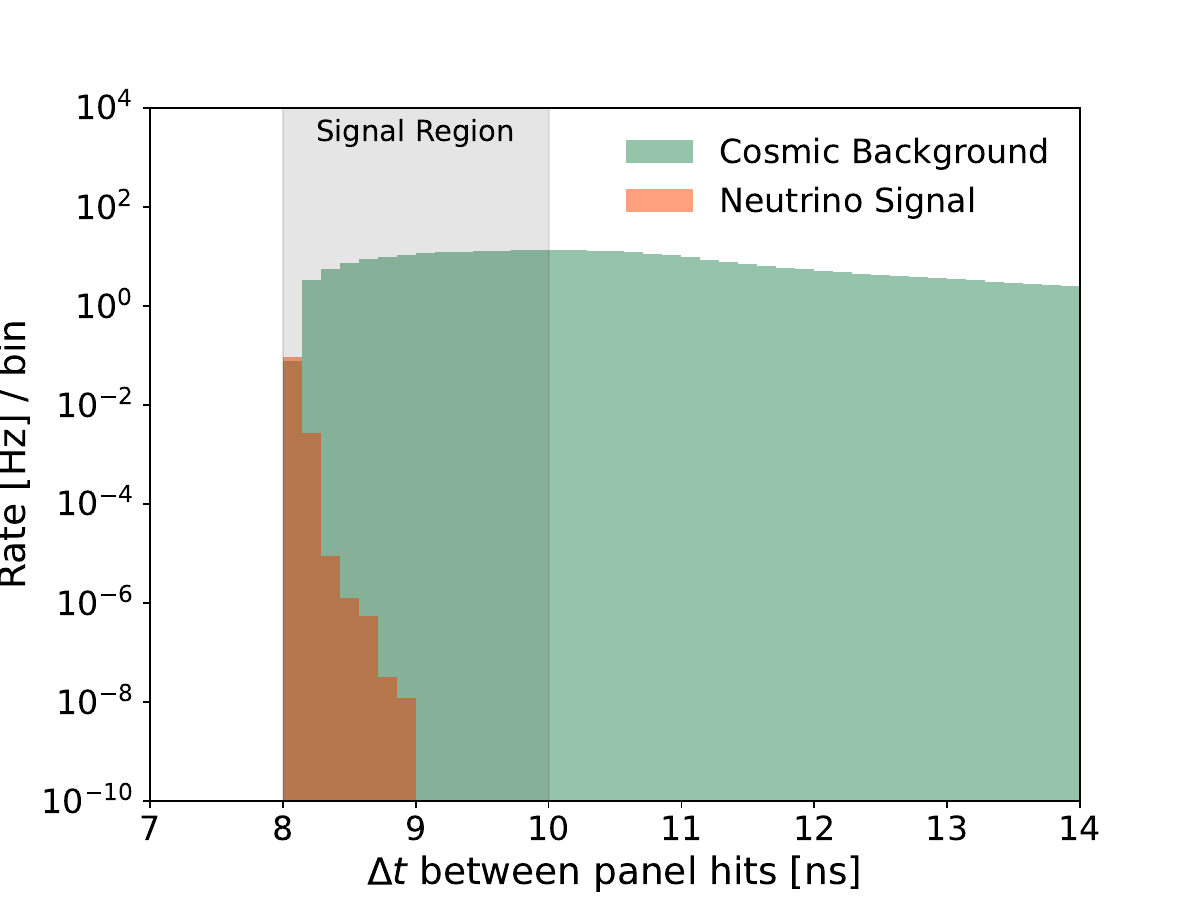}
    \caption{The distribution of $\Delta t$, the time difference between panel crossings in SINE, for signal and background muons. The proposed signal cut in indicated by the shaded region.}
    \label{fig:panel_deltaT}
\end{figure}

\begin{figure}
    \centering
    \includegraphics[width=\linewidth]{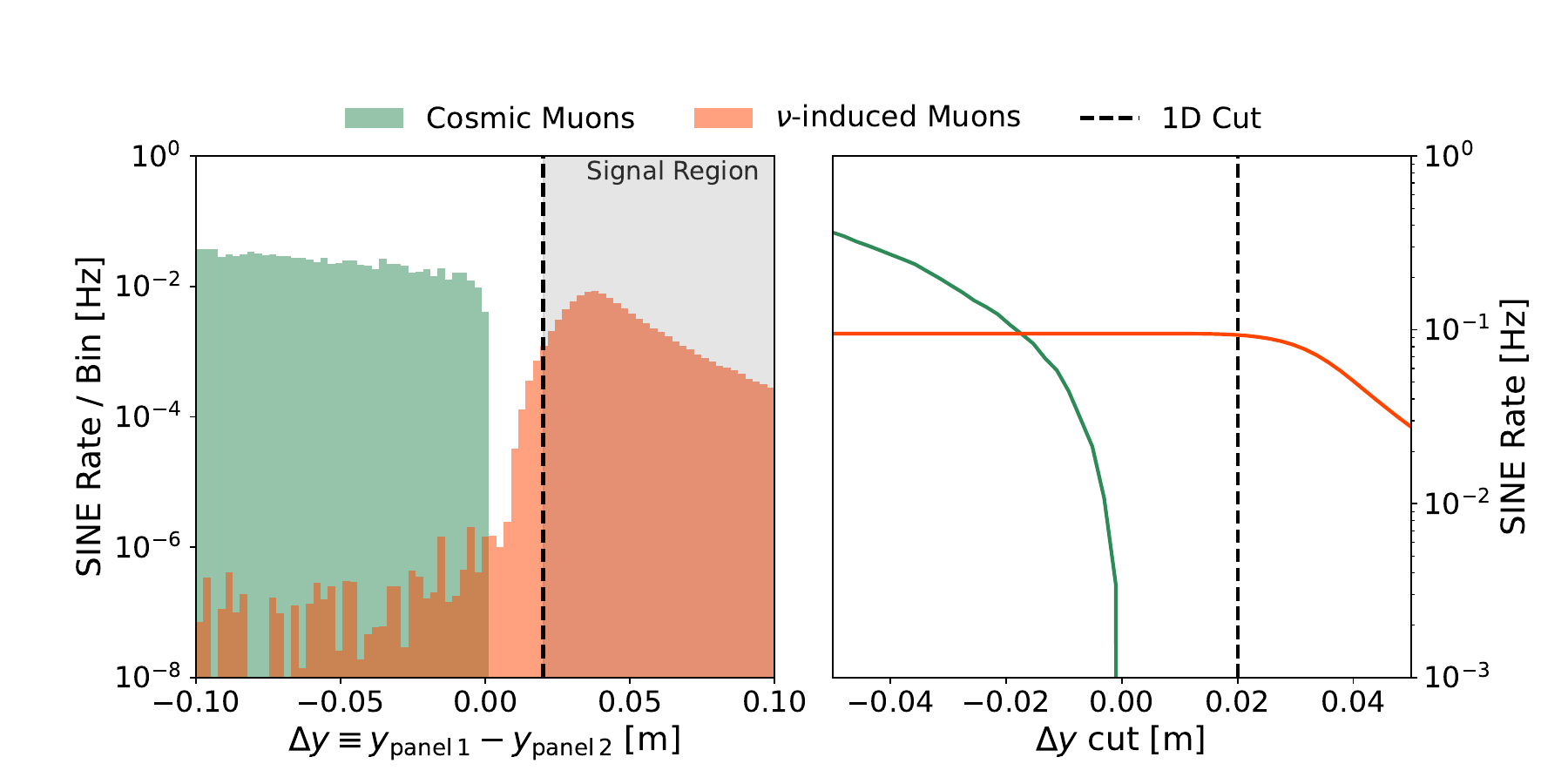}
    \caption{The left panel shows the distribution of $\Delta y$, the height difference between panel crossings in SINE, for signal and backgrounds, as well as the proposed signal cut in the shaded region. The right panel shows the result signal and background rates at SINE for different cuts on $\Delta y$, as well as the proposed $\Delta y > 2\,{\rm cm}$ cut.}
    \label{fig:vertical_cut}
\end{figure}

In this section, we provide further details on the proposed timing and spatial cuts described in \cref{sec:bkg}, designed to separate neutrino-induced muon signals from cosmic muon backgrounds in SINE.
\Cref{fig:beam_timing} shows the timing distribution of background muons in each SINE sub-detector.
Approximately $99.9\%$ of neutrino-induced muons arrive within a 2.5\,ns window, supporting the initial ``Beam Timing'' cut in \cref{fig:SINE_backgrounds}.
\Cref{fig:panel_deltaT} shows the distribution of $\Delta t$, the time difference between front and back scintillator panel crossings in SINE.
Signal muons are all contained within the signal region, $8 < \Delta t/{\rm ns} < 10$, while background muons are spread more uniformly in $\Delta t$, supporting the ``Panel $\Delta t$'' cut in \cref{fig:SINE_backgrounds}.
Finally, \cref{fig:vertical_cut} shows the distribution of $\Delta y$, the height difference between front and back scintillator panel crossings in SINE.
As expected, signal muons tend to result in $\Delta y>0$ while backgrounds muons tend to result in $\Delta y < 0$.
This motivates the displayed one-dimensional spatial cut $\Delta y > 2\,{\rm cm}$, supporting the ``1D Cut'' in \cref{fig:SINE_backgrounds}.

\subsection{Heavy Neutral Lepton Production}

\begin{figure}
    \centering
    \includegraphics[width=0.8\linewidth]{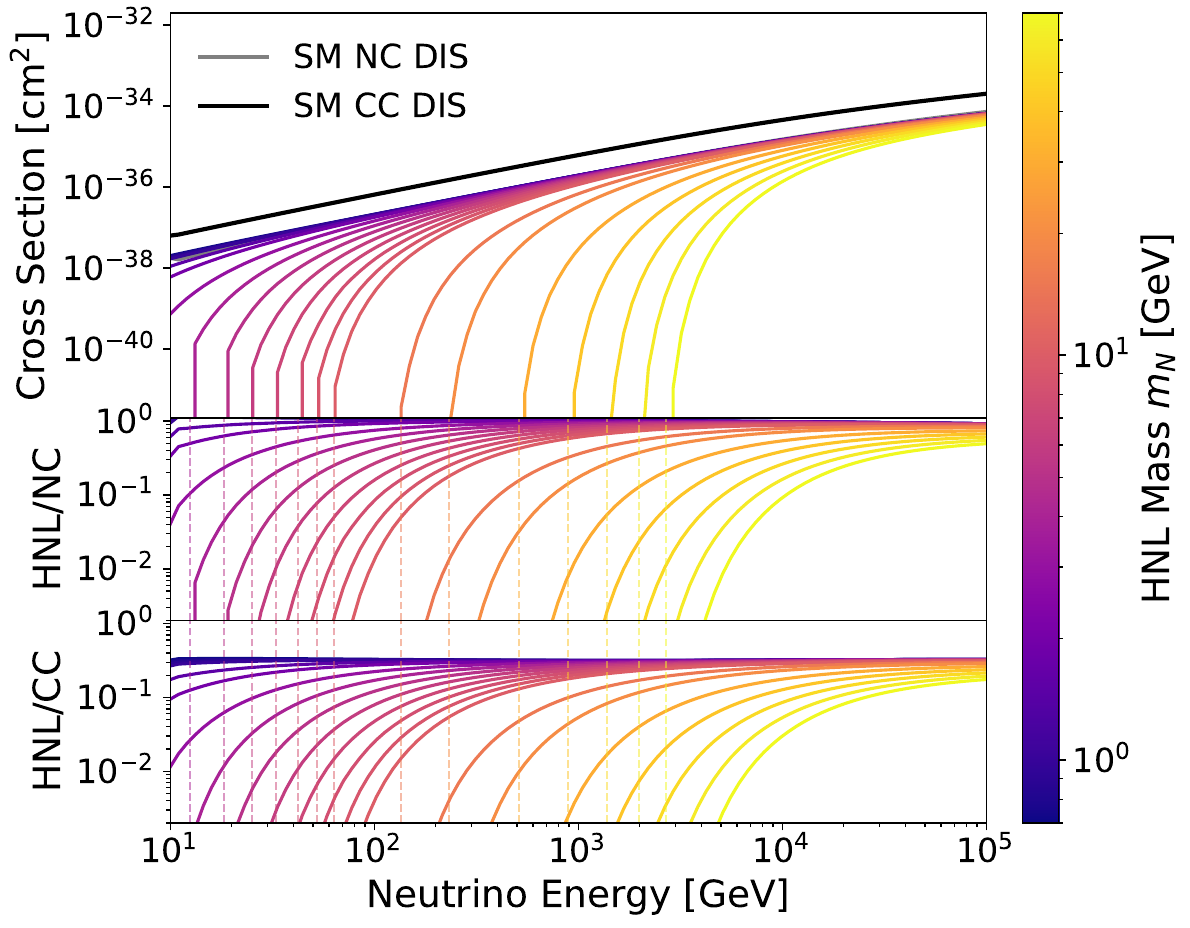}
    \caption{The HNL production cross sections used to compute the SINE sensitivities in \cref{fig:hnl}. The bottom panels show the ratio to the Standard Model CC and NC DIS cross sections from Ref.~\cite{Weigel:2024gzh}. Vertical dotted lines correspond to the threshold for each HNL mass. Cross sections are computed at unity mixing, $|U_{\mu N}|^2 = 1$.}
    \label{fig:hnl_xs}
\end{figure}

As discussed in \cref{sec:physics}, we consider HNL production via neutrino DIS when computing the sensitivities in \cref{fig:hnl}.
\Cref{fig:hnl_xs} shows the cross section for this process across the relevant range of HNL masses, as well as ratios to the Standard Model CC and NC DIS cross sections.
Here, one can see the impact of the kinematic threshold for different HNL masses.

\end{document}